\documentclass[12pt]{iopart}
\usepackage{iopams,color,graphicx,multirow}
\newcommand{\be}{\begin{equation}}
\newcommand{\ee}{\end{equation}}
\newcommand{\bea}{\begin{eqnarray}}
\newcommand{\eea}{\end{eqnarray}}
\newcommand{\um}{{\rm m}}
\newcommand{\ukm}{{\rm km}}
\newcommand{\ucm}{{\rm cm}}
\newcommand{\umm}{{\rm mm}}
\newcommand{\unm}{{\rm nm}}
\newcommand{\upm}{{\rm pm}}
\newcommand{\ufm}{{\rm fm}}
\newcommand{\uHz}{{\rm Hz}}
\newcommand{\ug}{{\rm g}}
\newcommand{\ukg}{{\rm kg}}

\newcommand{\umK}{{\rm mK}}
\newcommand{\umW}{{\rm mW}}
\newcommand{\us}{{\rm s}}
\newcommand{\uC}{{\rm C}}
\newcommand{\uV}{{\rm V}}
\newcommand{\kHz}{{\rm kHz}}
\newcommand{\mHz}{{\rm mHz}}
\newcommand{\mum}{\mu{\rm m}}
\newcommand{\muN}{\mu{\rm N}}
\newcommand{\muV}{\mu{\rm V}}

\newcommand{\TRC}{TianQin Research Center for Gravitational Physics \& School of Physics and Astronomy, Sun Yat-sen University (Zhuhai Campus), Zhuhai 519082, P.R. China}

\newcommand{\CGE}{Centre for Gravitational Experiments, School of Physics, MOE Key Laboratory of Fundamental Physical Quantities Measurement \& Hubei Key Laboratory of Gravitation and Quantum Physics, PGMF, Huazhong University of Science and Technology, Wuhan 430074, P. R. China}

\newcommand{\DFH}{DFH Satellite Co., Ltd., Beijing 100094, P.R. China}
\newcommand{\ICE}{Beijing Institute of Control Engineering, Beijing 100094, P.R. China}
\newcommand{\ISSE}{Beijing Institute of Spacecraft System Engineering, Beijing 100094, P.R. China}
\begin{document}

\title{The First Round Result from the TianQin-1 Satellite}

\author{Jun Luo$^{1,2}$, Yan-Zheng Bai$^{2}$, Lin Cai$^{2}$, Bin Cao$^1$, Wei-Ming Chen$^1$, Yu Chen$^1$, De-Cong Cheng$^2$, Yan-Wei Ding$^{1}$, Hui-Zong Duan$^{1}$, Xingyu Gou$^4$, Chao-Zheng Gu$^1$, De-Feng Gu$^{1}$, Zi-Qi He$^1$, Shuang Hu$^2$, Yuexin Hu$^{3}$, Xiang-Qing Huang$^{1}$, Qinghua Jiang$^4$, Yuan-Ze Jiang$^2$, Hong-Gang Li$^2$, Hong-Yin Li$^{2}$, Jia Li$^{2}$, Ming Li$^{3}$, Zhu Li$^{1}$, Zhu-Xi Li$^{2}$, Yu-Rong Liang$^{2}$, Fang-Jie Liao$^1$, Yan-Chong Liu$^2$, Li Liu$^{2}$, Pei-Bo Liu$^1$, Xuhui Liu$^4$, Yuan Liu$^{1}$, Xiong-Fei Lu$^1$, Yan Luo$^1$, Jianwei Mei$^{1}$, Min Ming$^{2}$, Shao-Bo Qu$^{2}$, Ding-Yin Tan$^{2}$, Mi Tang$^2$, Liang-Cheng Tu$^{1,2,a}$, Cheng-Rui Wang$^{2}$, Fengbin Wang$^3$, Guan-Fang Wang$^1$, Jian Wang$^1$, Lijiao Wang$^4$, Xudong Wang$^4$, Ran Wei$^5$, Shu-Chao Wu$^{2}$, Chun-Yu Xiao$^{2}$, Meng-Zhe Xie$^2$, Xiao-Shi Xu$^1$, Liang Yang$^1$, Ming-Lin Yang$^1$, Shan-Qing Yang$^{1}$, Hsien-Chi Yeh$^{1,b}$, Jian-Bo Yu$^{2}$, Lihua Zhang$^{3,c}$, Meng-Hao Zhao$^2$, Ze-Bing Zhou$^{2,d}$}
\address{$^1$\TRC}
\address{$^2$\CGE}
\address{$^3$\DFH}
\address{$^4$\ICE}
\address{$^5$\ISSE}
\ead{$^a$tuliangch@mail.sysu.edu.cn, $^b$yexianji@mail.sysu.edu.cn, $^c$zlh700717@sina.com, $^d$zhouzb@hust.edu.cn}
\vspace{10pt}
\begin{indented}
\item[Last updated:]~\today
\end{indented}

\newpage

\centerline{\bf Abstract}

The TianQin-1 satellite (TQ-1), which is the first technology demonstration satellite for the TianQin project, was launched on 20 December 2019. The first round of experiment had been carried out from 21 December 2019 until 1 April 2020. The residual acceleration of the satellite is found to be about $1\times10^{-10}~\um/\us^{2}/\uHz^{1/2}$ at $0.1~\uHz\,$ and about $5\times10^{-11}~\um/\us^{2}/\uHz^{1/2}$ at $0.05~\uHz\,$, measured by an inertial sensor with a sensitivity of $5\times10^{-12}~\um/\us^{2}/\uHz^{1/2}$ at $0.1~\uHz\,$. The micro-Newton thrusters has demonstrated a thrust resolution of $0.1~\muN$ and a thrust noise of $0.3~\muN/\uHz^{1/2}$ at $0.1~\uHz$. The residual noise of the satellite with drag-free control is $3\times10^{-9}~\um/\us^{2}/\uHz^{1/2}$ at $0.1~\uHz\,$. The noise level of the optical readout system is about $30~\upm/\uHz^{1/2}$ at $0.1~\uHz\,$. The temperature stability at temperature monitoring position is controlled to be about $\pm3~\umK$ per orbit, and the mismatch between the center-of-mass of the satellite and that of the test mass is measured with a precision of better than $0.1~\umm$.

%
%
%
%
%

\newpage
\tableofcontents

\section{Introduction}\label{sec:intro}

The TianQin project was initiated in 2014. The goal is to launch the space-based gravitational-wave (GW) observatory TianQin around 2035 and to detect GWs in the frequency range $10^{-4}\sim1~\uHz$ \cite{Luo:2015ght}.
The major sources of TianQin include the inspiral of ultra-compact Galactic binaries , the inspiral of stellar mass black holes within a redshift of the order $z\sim0.1$, the extreme mass ratio inspirals within a redshift of the order $z\sim1$, the merger of massive black holes at cosmological distances (with any redshift given that there is a source), and possibly also GWs from the very early universe or from exotic sources such as cosmic strings \cite{Hu:2017yoc,Liu:2020,Wang:2019ryf}.
With the detection of such sources, TianQin is expected to provide key information on the astrophysical history of galaxies and black holes, the dynamics of dense star clusters and galactic centers, the nature of gravity and black holes, the expansion of the universe, and possibly also the fundamental physics related to the phase transition of the universe \cite{Shi:2019hqa,Bao:2019kgt}.

TianQin consists of three satellites on nearly identical geocentric orbits with radii of the order $10^5~\ukm$, forming a normal triangle constellation.
There are test masses (TMs) in each satellite and the satellites are drag-free controlled to precisely follow the motion of the TMs along the sensitive axes, so that the nongravitational disturbance is minimised.
Laser interferometry is utilized to measure the variation in the optical path length between the TMs.
In order to achieve the scientific goals, the nongravitational disturbance on the TMs must be reduced to the order of $10^{-15}~\um/\us^{2}/\uHz^{1/2}$ in $10^{-4}\sim1~\uHz$ band, and the noise of the displacement measurement with laser interferometry must be reduced to the order of $10^{-12}~\um/\uHz^{1/2}$ in $10^{-4}\sim1~\uHz$ band \cite{Luo:2015ght}. These are the two core technology requirements for TianQin.

The research of high precision space inertial sensors dates to the 1950s, when they were used to monitor the disturbance of satellites in space environment investigation missions.
The technology of geodesic tracking with the guidance of inertial sensing was originally proposed at Stanford University for the TRIAD satellite in the 1960s \cite{DeBra1973,JHUHU1974,1976JGR....81.3753M}.
ONERA has successfully used cubic TMs and six degrees of freedom electrostatic servo-control in several geodesy missions including the CHAMP, GRACE, GOCE, MicroSCOPE, and GRACE Follow-On missions  \cite{Touboul:20338,Touboul:2017grn,Abich:2019cci}.
In these missions, the inertial sensors were working in the accelerometer mode measuring non-gravitational forces on low-orbit satellites.
When working in the inertial reference mode, the inertial sensor can be used as the inertial reference for space-based GW detection missions \cite{Dolesi_2003}.
On 3 December 2015, the technology demonstration spacecraft LISA Pathfinder was successfully launched.
The results of LISA Pathfinder indicate that the performance of its inertial sensors has met the requirement of LISA \cite{Armano:2016bkm}.

The study of ultra-stable laser interferomter for space-based GW mission started since the last decades of the twentieth century \cite{Hirth2009,Audley:2011zz,Killow:16,Robertson:18}.
LISA Pathfinder uses an ultra-precision laser interferometer equiped with a quasi-monolithic optical bench to detect the relative motion between the two TMs onboard the spacecraft.
The interferomter accomplished $34.8~\ufm/\uHz^{1/2}$ above $60~\mHz$, for a light-path baseline of about $38~\ucm$ \cite{Armano:2016bkm}.
The GRACE Follow-On mission, launched on 19 May 2018, carries a laser ranging system that provides the first inter-satellite laser interferometric ranging between remote spacecraft.
The laser ranging system has achieved a noise level of $1~\unm/\uHz^{1/2}$ above $0.1~\uHz$, over a distance of about $220~\ukm$ \cite{Abich:2019cci}.

In order to bring the key technologies of TianQin to mature, a technology roadmap called the 0123 plan has been adopted in 2015:
\begin{itemize}
\item Step 0: Acquiring the capability to obtain high precision orbit information for satellites in the TianQin orbit through lunar laser ranging experiments. This step involves constructing new laser ranging stations on the ground and creating a new generation of corner-cube retro-reflectors (CCR) to be installed on the targets.
\item Step 1: Using single satellite missions, and the main goal is to test and demonstrate the maturity of the inertial reference technology;
\item Step 2: Using a mission with a pair of satellites, and the main goal is to test and demonstrate the maturity of the inter-satellite laser interferometry technology;
\item Step 3: Launching a triple of satellites to form the space-based GW observatory, TianQin.
\end{itemize}
Projects in step 0 have started since 2016. A $17-\ucm$ single hollow CCR has been created and launched with the Chang'E 4 relay satellite, QueQiao, on 21 May 2018 \cite{HeLiu:2018}. A new laser ranging station equipped with a $1.2-\um$ telescope has been constructed on the top of the Fenghuang mountain by the Zhuhai Campus of Sun Yat-sen University. This station has become the first in China to successfully receive laser ranging signals echoed from all the five retro-reflectors on the Moon and the result will be reported in a separate paper.

TianQin-1 (TQ-1 for short) is a mission in Step 1.
The major objectives of TQ-1 include testing the technologies of inertial sensing, micro-Newton propulsion, drag-free control and laser interferometry with in-orbit experiments.
Additional objectives of TQ-1 include testing the technologies of temperature control and center-of-mass (CoM) measurement of the satellite.
The preparation for TQ-1 started in 2016 and the project was officially approved by the China National Space Administration (CNSA) in 2018.

TQ-1 was successfully launched on 20 December 2019 from the Taiyuan Satellite Launch Center in Shanxi Province.
The satellite completed its startup phase on 21 December 2019 and has been functioning smoothly since after.
In this paper, we report the results from the first round of experiments carried out from 21 December 2019 until 1 April 2020.

The TQ-1 has achieved the following performances:
\begin{enumerate}
\item The inertial sensor, which has been calibrated to have a sensitivity of $5\times10^{-12}~\um /\us^{2}/ \uHz^{1/2}$ at $0.1~\uHz$ on the ground, measured the residual acceleration of the satellite in flight to be about $1\times10^{-10}~\um/\us^{2}/\uHz^{1/2}$ at $0.1~\uHz$ and about $5\times10^{-11}~\um/\us^{2}/\uHz^{1/2}$ at $0.05~\uHz$;
\item The micro-Newton thrusters demonstrated a thrust resolution of about $0.1~\muN$ and thrust noise of about $0.3~\muN/\uHz^{1/2}$ at $0.1~\uHz$;
\item The residual noise of the satellite with drag free control is about $3\times10^{-9}~\um/\us^{2}/\uHz^{1/2}$ at $0.1~\uHz\,$;
\item The noise level of the optical readout system is about $30~\upm/\uHz^{1/2}$ at $0.1~\uHz\,$;
\item The temperature at key temperature monitoring position has been controlled to be about $\pm3~\umK$ per orbit;
\item The mismatch between the CoM of the satellite and that of the TM has been measured to be about $0.1~\umm$.
\end{enumerate}

The paper is organised as follows. In section \ref{sec:TQ-1}, we outline the basic design and the experiment principles of TQ-1. In section \ref{sec:rst}, we present the results of the first round experiments. In section \ref{sec:discussion}, we conclude by discussing how the results obtained with TQ-1 may impact the development of the TianQin project and the development of space-based GW detection as a whole.

\section{The satellite}\label{sec:TQ-1}

TQ-1 is a low cost microsatellite with a mass of about $103~\ukg$. The satellite uses a sun-synchronous orbit with an altitude of $628~\ukm$, an inclination angle of $98^\circ$ and an orbital period of $97.13\min$. The body of TQ-1 is in the shape of a cuboid with three sides measuring $670~\umm~(X)\times 670~\umm~(Y) \times644.5~\umm~(Z)$, where $X$ is the flight direction and $Z$ is the direction pointing to the center of the Earth.

\begin{figure}
\includegraphics[width=\linewidth]{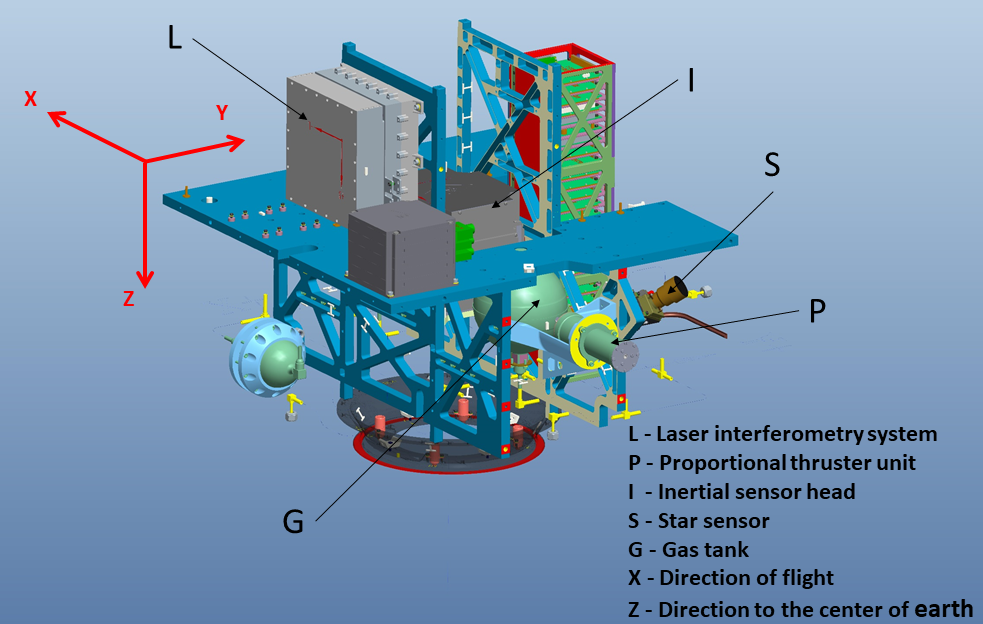}
\caption{The installation of payloads inside the TQ-1 satellite.}
\label{fig:TQ-1}
\end{figure}

A picture of TQ-1 is given in Fig. \ref{fig:TQ-1}.
At the center of the satellite is the inertial sensor head, which contains a TM measuring $10~\umm \times40~\umm \times40~\umm$ in size.
The mismatch between the CoM of the satellite and that of the TM has been adjusted and reduced to be less than $0.1~\umm$ before launch.
There is a pair of cold gas proportional micro-thrusters with the line of force pointing along the flight direction of the satellite.
The laser interferometry system is separated from the inertial sensing system.
There is active and high precision temperature control during experiments of the inertial sensor and the laser interferometry system.

\section{The result}\label{sec:rst}

The main types of experiments carried out between 21 December 2019 and 1 April 2020 are listed in Table \ref{tab:exps}. In this section, we present the main results from each of these experiments.

\begin{table}
\caption{Main types of experiments carried out on TQ-1. A ``$\surd$" mark means the corresponding system was turned on. Acronyms: Inertial sensor (IS), micro-Newton thruster (MT), drag free control (DFC), and laser interferometry (LI).}
\label{tab:exps}
\begin{tabular}{|c|c|c|c|c|c|}
  \hline
  Experiment type & Key technology to test & IS & MT & DFC & LI \\
  \hline
  {\bf Type-1} & Inertial sensing & $\surd$ &  &  & \\
  \hline
  {\bf Type-2} & Micro-propulsion & $\surd$ & $\surd$ &  & \\
  \hline
  {\bf Type-3} & Dragfree Control & $\surd$ & $\surd$ & $\surd$ & \\
  \hline
  {\bf Type-4} & Laser interferometry &  &  &  & $\surd$ \\
  \hline
\end{tabular}
\end{table}

\subsection{The inertial sensing system}

\begin{figure}
\includegraphics[width=\linewidth]{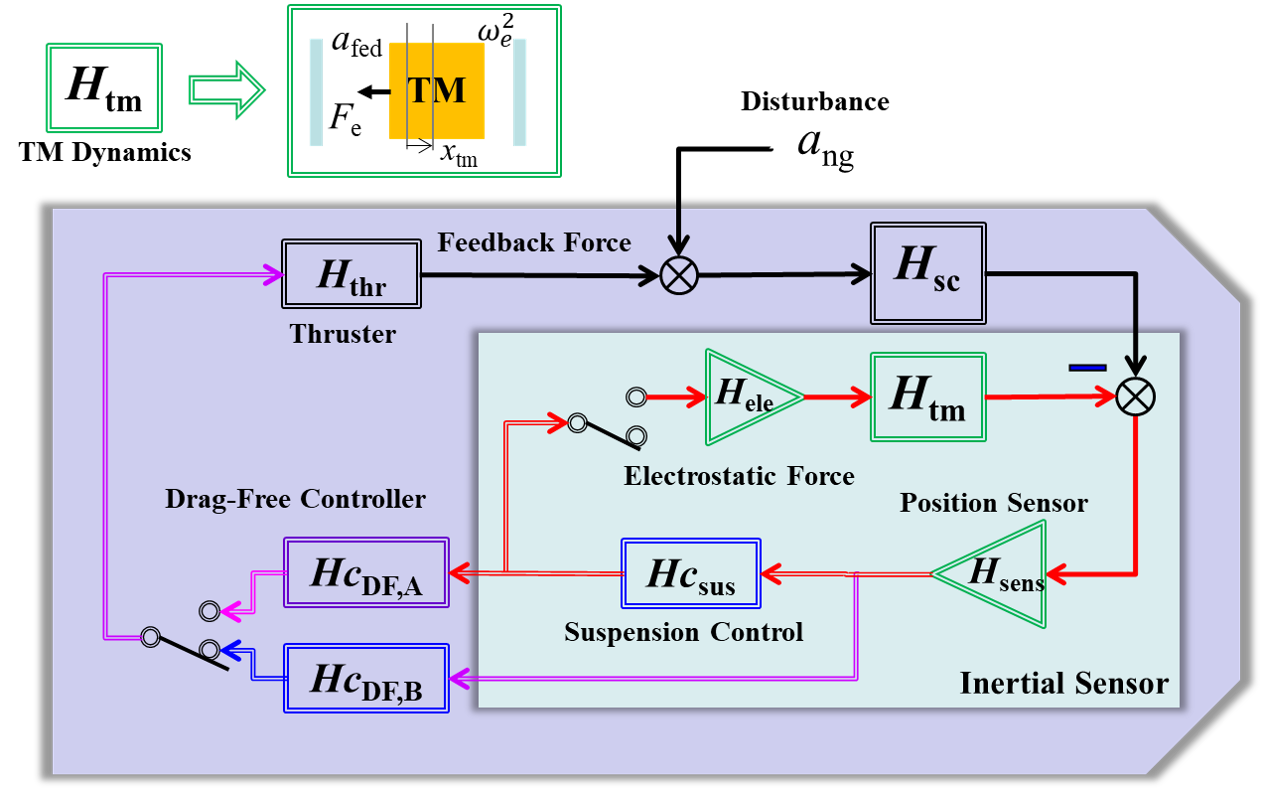}
\caption{Principle of the inertial sensing and the drag-free control system for TQ-1. $H_{tm}$ is the dynamics of TM, describing its motion response to the input acceleration; $F_e/a_{fed}$ is the suspension control electrostatic force and acceleration; $x_{tm}$  is the linear displacement of the TM; $H_{sens}$ is the transfer function of capacitive displacement sensor; $H_{ele}$ is the electrostatic actuator transfer function; $Hc_{sus}$ is the test mass suspension controller; $Hc_{DF,A}$ is the drag-free controller in accelerometer mode; $Hc_{DF,B}$ is the drag-free controller in TM free fly mode; $H_{thr}$ is the transfer function of micro-Newton thruster; $H_{SC}$ is the dynamics of the satellite.}
\label{fig:IS-DFC}
\end{figure}

The inertial sensing system of TQ-1 is based on capacitive sensing and electrostatic control.
The study and development of such sensors have started since 2000 at the Centre for Gravitational Experiments, Huazhong University of Science and Technology (HUST-CGE).
Several fight models have been constructed and successively tested in orbit \cite{s17091943}.

The core probe of the inertial sensing system consists of one test mass (TM) and capacitive displacement sensors surrounding the TM.
As the floating TM and the frame connected to the satellite are subjected to different forces, the displacements between them will vary and then are monitored by the sensors.

The drag-free control of TQ-1 is performed only along the flight direction of the satellite. As the inertial sensor outputs the acceleration signal when the TM is controlled at the balanced position within the inertial sensor. Alternatively, the inertial sensor outputs the displacement signal when the TM is in a free-fall motion without electrostatic control in the flight direction of the satellite. The former operation mode of the inertial sensor is called the acceleration mode, and the latter is called the inertial reference mode. In the first round of experiment of TQ-1, the inertial sensor is operating with the acceleration mode. The acceleration signal from the inertial sensor is fed to the drag-free controller, and the micro-thrusters generate a force opposite to the direction of the external disturbance in the flight direction of the satellite until the acceleration measured by the inertial sensor is reduced to a minimum that is limited by the thrust noise of the micro-Newton thrusters.
The principle of the inertial sensing system and the drag-free control loop onboard of TQ-1 is given in Fig. \ref{fig:IS-DFC}.

In the accelerometer mode, the sensor output is sent to the suspension controller $Hc_{sus}$, which sends a signal commanding the electrostatic actuator $H_{ele}$ to apply an accurate servo-acceleration on the TM to pull it to the balance position. The control signal from $Hc_{sus}$ is also sent to the drag-free controller $Hc_{DF,A}$, which commands the micro-Newton thrusters $H_{thr}$ to minimise the measured non-gravitational acceleration $a_{n}$.

In the inertial reference mode, the suspension control is shut to minimize direct disturbances acting on the TM, which is expected to be adopted along the sensitive axis for GW detection missions, and the output of the displacement sensor is sent to $Hc_{DF,B}$, which then commands the thrusters $H_{thr}$ to compensate the non-gravitational acceleration $a_{n}$.

\begin{figure}
\includegraphics[width=\linewidth]{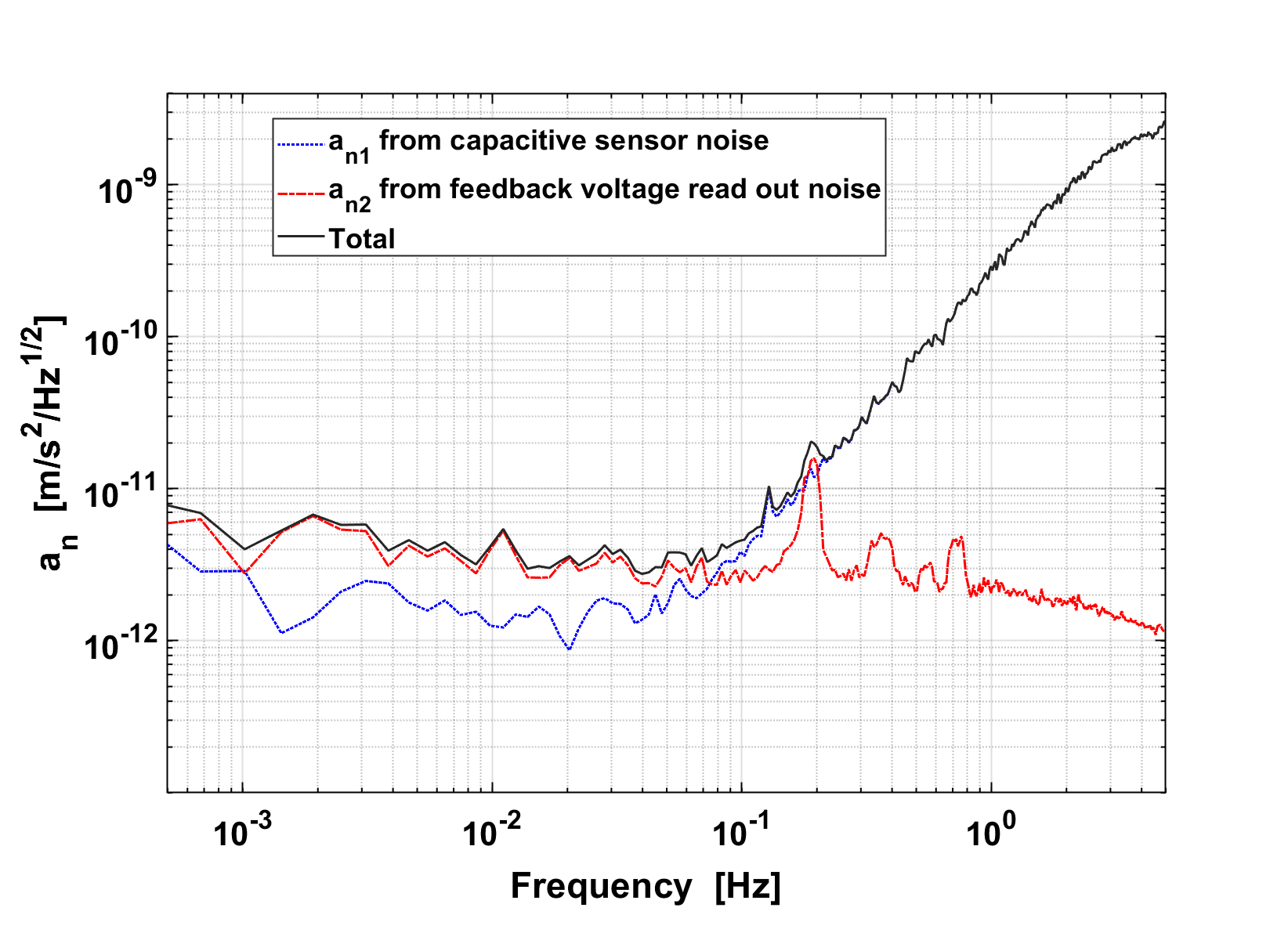}
\caption{The noise level of the inertial sensor. The red and blue curves are the contributions from the voltage readout noise and the capacitive displacement sensor noise, respectively. The black curve is the total noise.}
\label{fig:IS}
\end{figure}

\begin{figure}
\includegraphics[width=\linewidth]{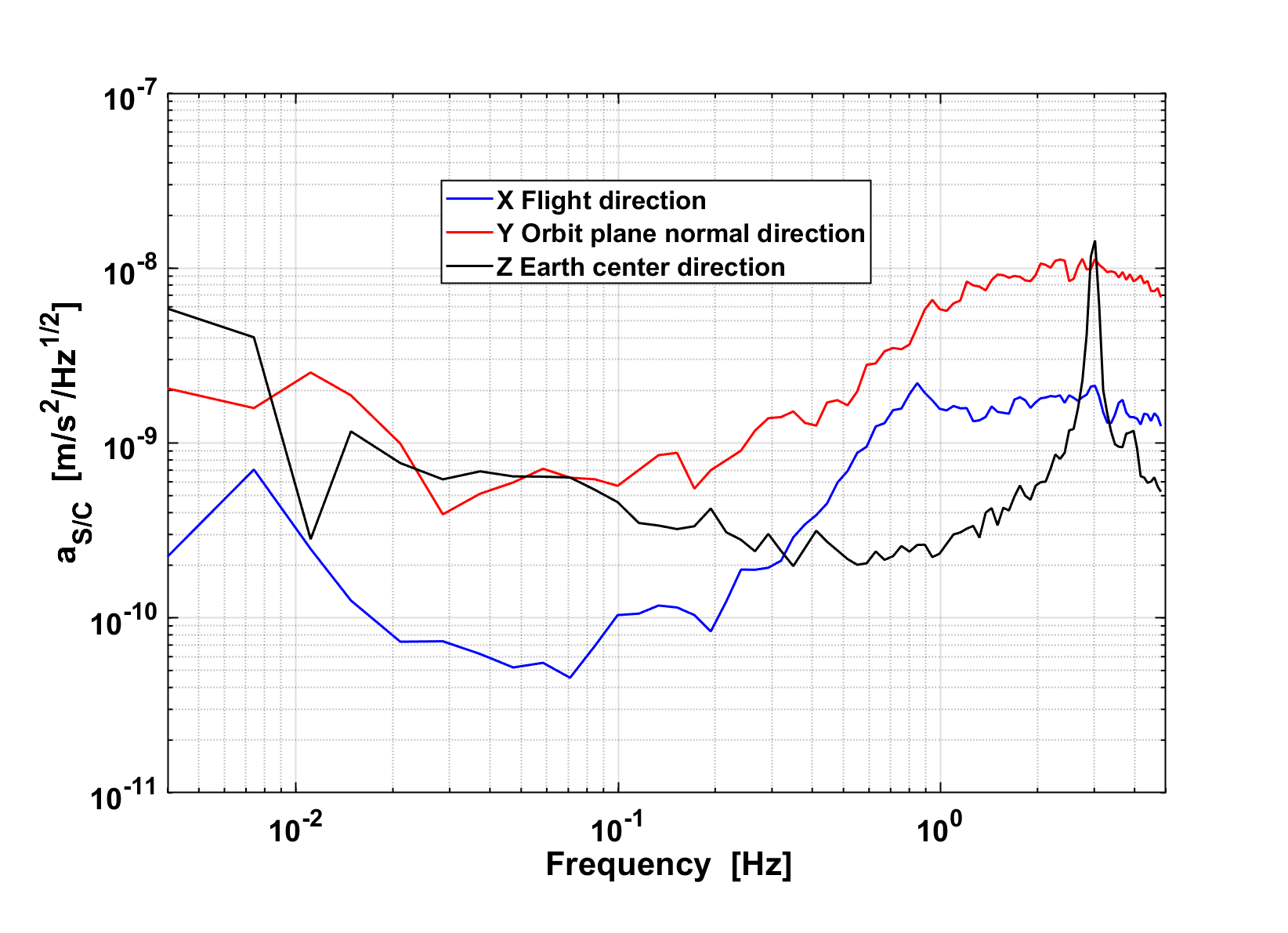}
\caption{The actual acceleration level of the satellite measured by the inertial sensor onboard TQ-1. Note $X$ is along the flight direction and $Z$ points to the center of Earth.}
\label{fig:IS2}
\end{figure}

The sensitivity of the inertial sensor in the acceleration mode is given by
\be a_{n}=(w^2+w_e^2)x_n+V_nH_a\,,\label{acc.noise}\ee
where $w$ is the signal frequency, $w_e$ is the coefficient of electrostatic stiffness mainly due to the electrostatic suspension, $x_n$ is the measurement noise of the capacitive displacement sensor, $V_n$ is the output voltage noise of the electrostatic actuator, and $H_a$ is the transfer function of the electrostatic control unit. For the inertial sensor on TQ-1, a $72~\ug$ cubic Ti-alloy TM is chosen based on mature experiences \cite{s17091943} with a capacitive gap of about $145 ~\mum$ along the high-sensitivity axes.
The capacitive sensing and voltage readout noises are measured based on the actual electric circuits, where $x_n$ is about $10~\upm/\uHz^{1/2}$ and $V_n$ is about $2~\muV/\uHz^{1/2}$ at $0.1$ Hz \cite{doi:10.1063/1.4873334,doi:10.1063/1.4749845}.
The sensitivity and dynamic range of the electrostatic actuator are $(1.05\pm0.03)\times10^{-6}~\um/\us^{2}/\uV$ and $(-1\sim1)\times 10^{-5}~\um/\us^{2}$, respectively, which are calibrated by fiber suspension and high-voltage levitation methods \cite{Tu_2010,doi:10.1063/1.4833398}.
Based on Eq. (\ref{acc.noise}), the sensitivity of the inertial sensor is estimated to be approximately $5\times10^{-12}~\um/\us^{2}/\uHz^{1/2}$ at $0.1~\uHz$, as is shown by the black curve in Fig. \ref{fig:IS}, which is limited by the voltage readout noise at low frequencies and the displacement sensing noise at high frequencies, respectively.

Due to the solar pressure and the atmosphere drag effect, the overall non-gravitational acceleration noise of the satellite is at the level of $10^{-7}~\um/\us^{2}$. Owing to the precise installation and the stability of the center of the inertial sensor with the satellite, actual acceleration noise measured along the flight velocity of the satellite in a {\bf Type-1} experiment is about $10^{-10}~\um/\us^{2}/\uHz^{1/2}$ at $0.1~\uHz$ and $5\times10^{-11}~\um/\us^{2}/\uHz^{1/2}$ at $0.05~\uHz$, as is shown by the blue curve in Fig. \ref{fig:IS2}, which is mainly limited by attitude coupling, non-gravitational force variation, structure vibration of the satellite and so on. It is noted that the result reported here is preliminary and without any post data processing. Much more data accumulation and further detailed analysis are being performed, and then more detailed results will be released.

\subsection{The micro-Newton thrusters}

\begin{figure}
\includegraphics[width=\linewidth]{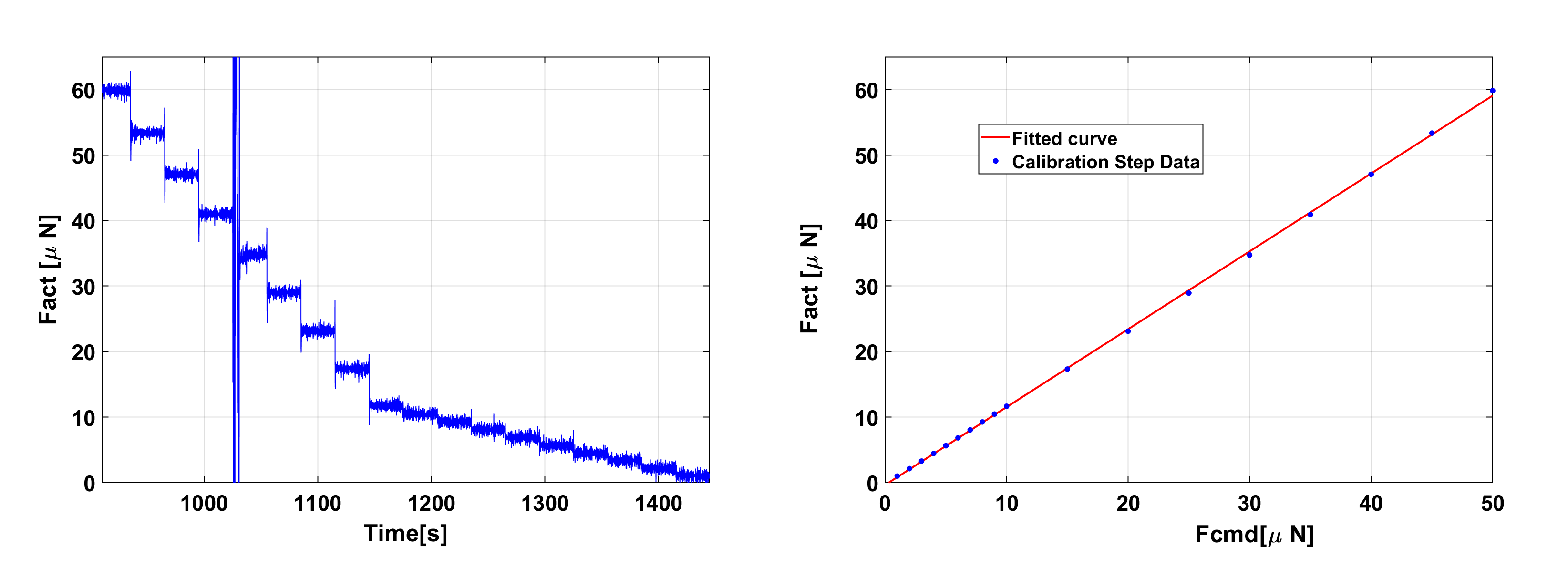}
\caption{The in-orbit calibration for the micro-Newton cold-gas thruster by the inertial sensor. (Left) Dynamic range of the micro-Newton thruster from $1\sim60~\muN$, measured by the inertial sensor. (Right)  Linear fitting of the command force $F_{cmd}$ and the actual force $F_{act}$, with the corresponding formula given in Eq. \ref{eq:fit}.}
\label{fig:MT2}
\end{figure}

The cold-gas micro-Newton thrusters onboard TQ-1 have a force resolution of about $0.1~\muN$.
In a {\bf Type-2} experiment, a dynamic range of $1\sim60~\muN$ has been tested, and the result is given in Fig. \ref{fig:MT2} (left).
The relation between the command force $F_{cmd}$ of the micro-Newton propulsion system and the actual force $F_{act}$ acting on the satellite measured by the inertial sensor is fit, as shown in Fig. \ref{fig:MT2} (right).
The sensitivity of the cold-gas micro-thruster in orbit is calibrated to be
\be {F_{act}} =(1.19 \pm 0.04)F_{cmd}-(0.35\pm0.12)~\muN\,.\label{eq:fit}\ee
For periods when the output of the micro-Newton thruster is kept constant, the overall noise of the thrust is found to be $0.3~\muN/\uHz^{1/2}$ at $0.1~\uHz$.

\subsection{Preliminary verification of the drag free control}

\begin{figure}
\includegraphics[width=\linewidth]{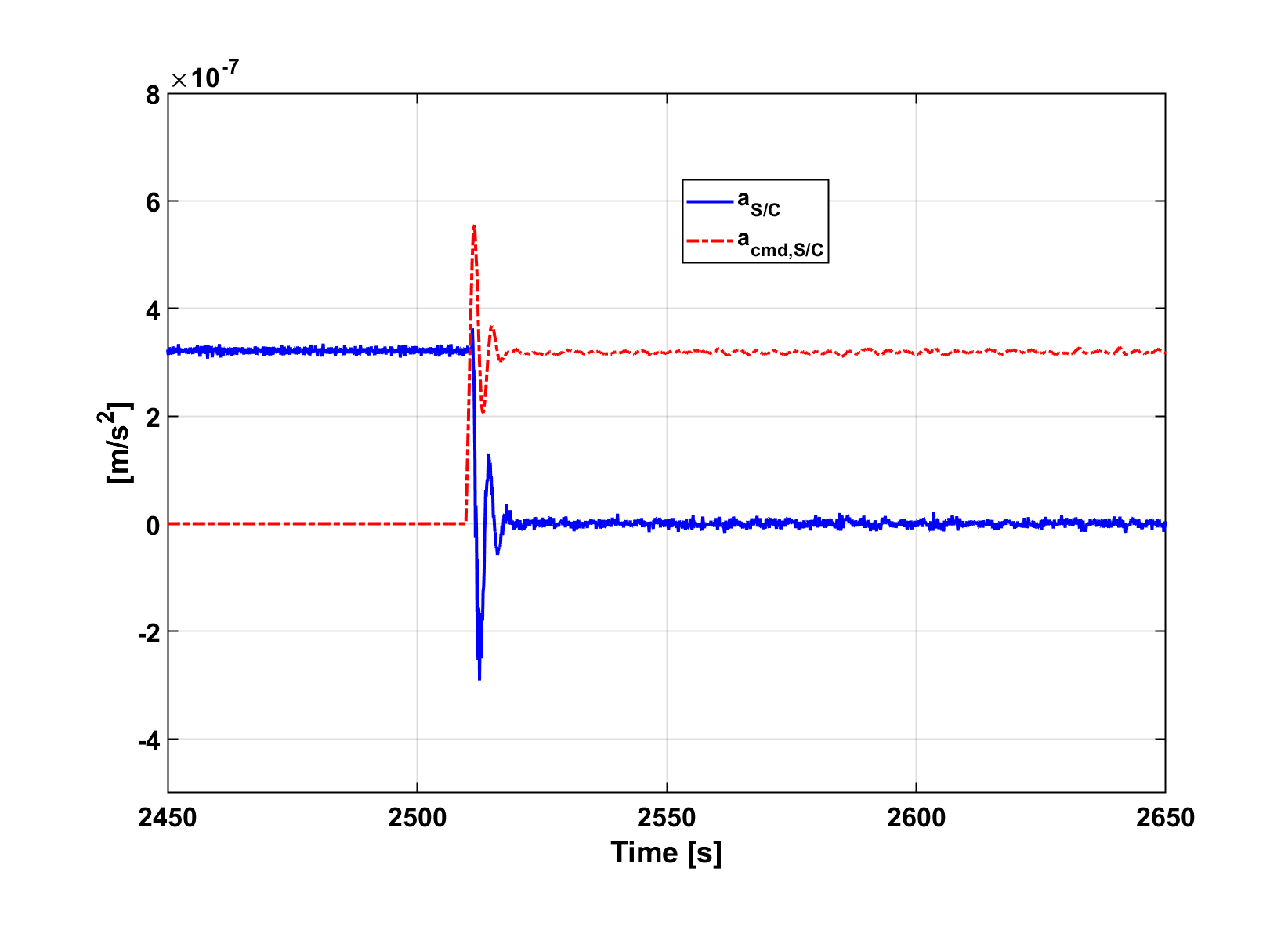}
\caption{In-orbit verification of the drag-free control of TQ-1. The solid-blue curve is the acceleration of the satellite measured by the inertial sensor, while the dashed-red curve is the command force sent to the thrusters by the drag-free controller to minimise the acceleration of the satellite.}
\label{fig:DFC}
\end{figure}

\begin{figure}
\includegraphics[width=\linewidth]{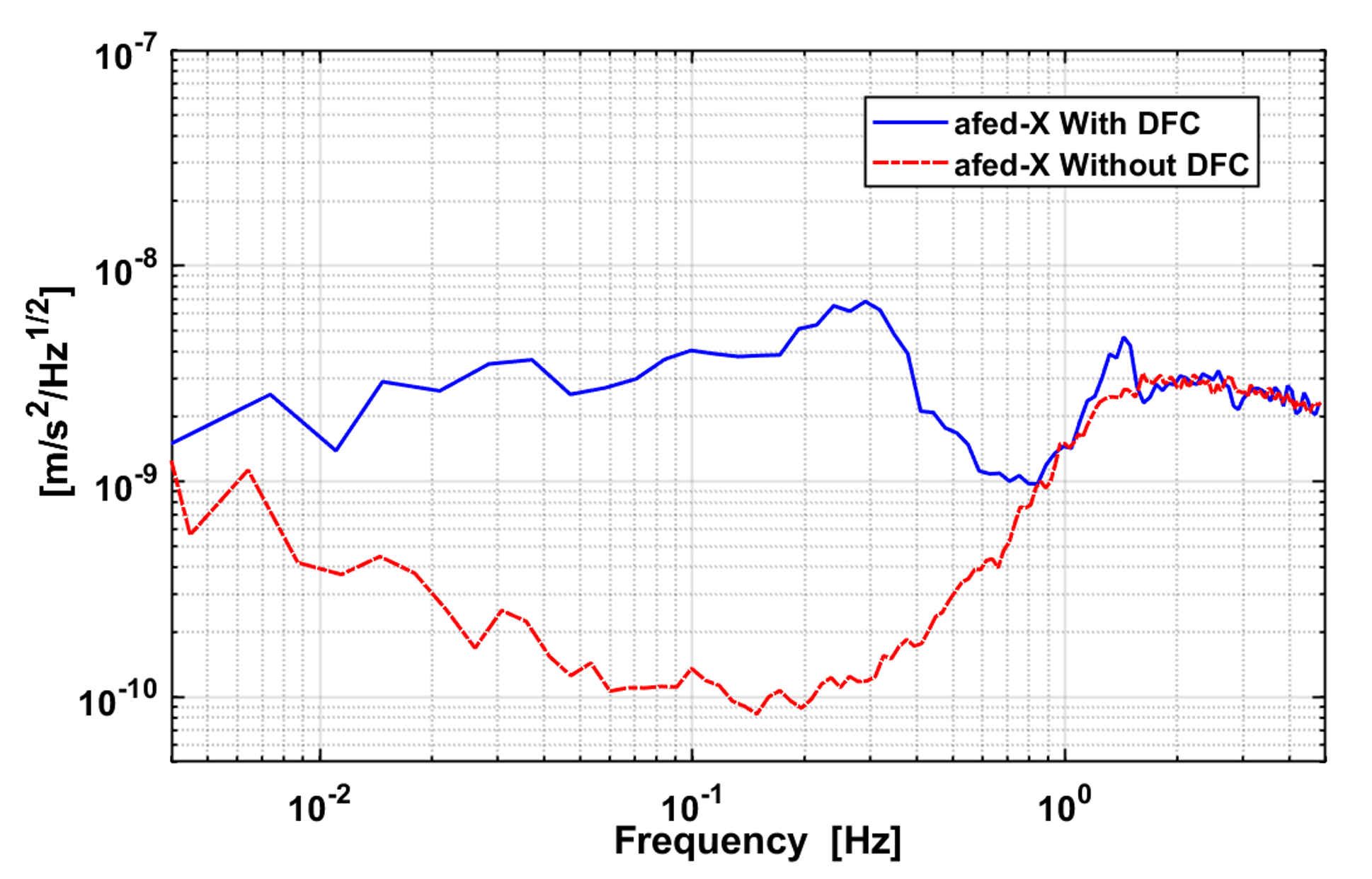}
\caption{The acceleration noise spectrums read from the inertial sensor with and without applying the drag-free control along the flight direction of TQ-1.}
\label{fig:DFC2}
\end{figure}

The drag-free control along the flight direction has been tested in the first round of experiment in orbit. In a {\bf Type-3} experiment with the acceleration measurement mode, the overall acceleration level of the satellite was reduced from $3.1\times10^{-7}~\um/\us^{2}$ to a nearly negligible level in less than $30~\us$ after the drag free control was switched on, as shown in Fig. \ref{fig:DFC}. The residual acceleration noise of the satellite is about $3\times10^{-9}~\um/\us^{2}/\uHz^{1/2}$ near $0.1~\uHz$ under the drag-free control condition. It is noted that the noise is increased by about 30 times compared to that without the drag-free control, which is caused by the thrust noise of about $0.3~\muN/\uHz^{1/2}$ at $0.1~\uHz$ and a control bandwidth of $0.07~\uHz$ (see Fig. \ref{fig:DFC2}). In next step, we will tune the drag-free controller to achieve a quicker response and improve the performance of the micro-Newton thrusters to achieve a lower thrust noise at low frequencies.

\subsection{The optical readout system}

\begin{figure}
\includegraphics[width=\linewidth]{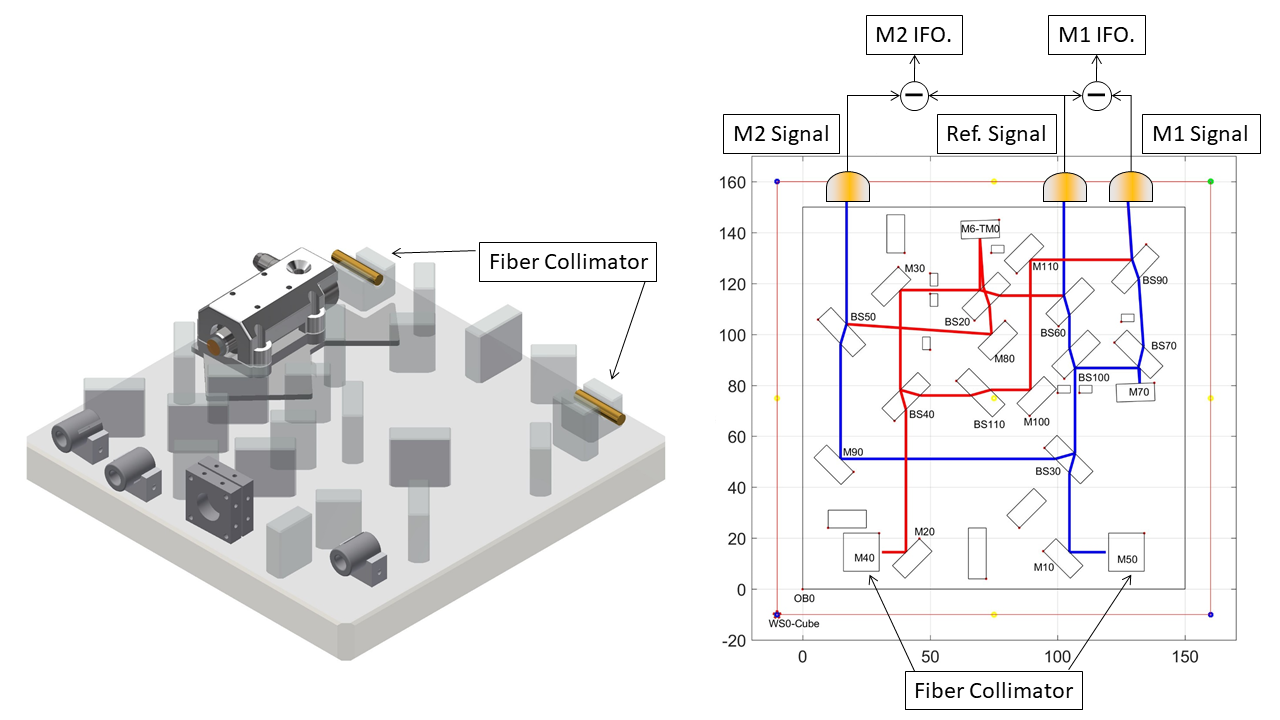}
\caption{A CAD model of the optical bench (left) and the layout of the optical path (right). M6-TM0 is the flexure hinge and M70 is the PZT-driven mirror. The laser beams are directed into the optical bench through two collimators (M40 and M50). After passing through different optical paths, the laser beams are recombined with BS50, BS60 and BS90 to generate three heterodyne signals. The optical paths of the reference signal: red-beam (M40-M20-BS40-M30-BS20-BS60) and blue-beam (M50-M10-BS30-BS100-BS60). The optical paths of M1 singal: red-beam (M40-M20-BS40-BS110-M100-M110-BS90) and blue-beam (M50-M10-BS30-BS100-BS70-M70-BS70-BS90). The optical paths of M2 singal: red-beam (M40-M20-BS40-M30-BS20-M6-BS20-M80-BS50) and blue-beam (M50-M10-BS30-M90-BS50).}
\label{fig:ORS}
\end{figure}

TQ-1 carries an Optical Readout System (ORS), which is a heterodyne laser interferometry system consisting of a heterodyne laser unit, an optical bench interferometer (OBI) and a phasemeter.
The heterodyne laser unit includes a Distributed-Bragg-Reflector (DBR) laser head and a pair of acousto-optic modulators (AOMs).
The wavelength of the DBR laser is $1064~\unm$ and the output power is $20~\umW$.
The two laser beams from the DBR laser are frequency-shifted with a $10-\kHz$ frequency offset by using AOMs and then guided into the OBI.
All of the beam connections between the laser head, the AOMs and the OBI are performed with single-mode polarization maintaining fibers.
The optical bench, composed of ULE substrate and fused-silica-based optical components, is assembled with the silicate bonding technique \cite{Elliffe_2005}.
A CAD model of the optical bench and the layout of the optical path are shown in Fig. \ref{fig:ORS}.

On the OBI, there are three interference beat-note signals detected by three single-element photodetectors. As shown in Fig. \ref{fig:ORS}, the middle beat-note signal is the reference, which combines with the right beat-note signal to form the measurement signal of the first Mach-Zehnder interferometer, named M1 interferometer. The reference signal combines with the left beat-note signal to form the measurement signal of the second Mach-Zehnder interferometer, named M2 interferometer.
Finally the relative phases of the signals are measured by the phasemeter. The working principle of the phasemeter is based on the digital phase-locked loop that is realized on an FPGA platform.

The M1 interferometer is used to measure the motion of a PZT-driven mirror (M70), and the M2 interferometer is used to measure the motion of a flexure hinge (M6-TM0) with a fixed end attached on the optical bench.
The PZT, mounted on the optical bench, is used to generate the displacements for calibration.
The function of the flexure hinge is to simulate a test mass.
The contrasts of the interference signals for the reference, the PZT-driven mirror and the flexure hinge are 0.7, 0.76, 0.81, respectively. 

\begin{figure}
\includegraphics[width=\linewidth]{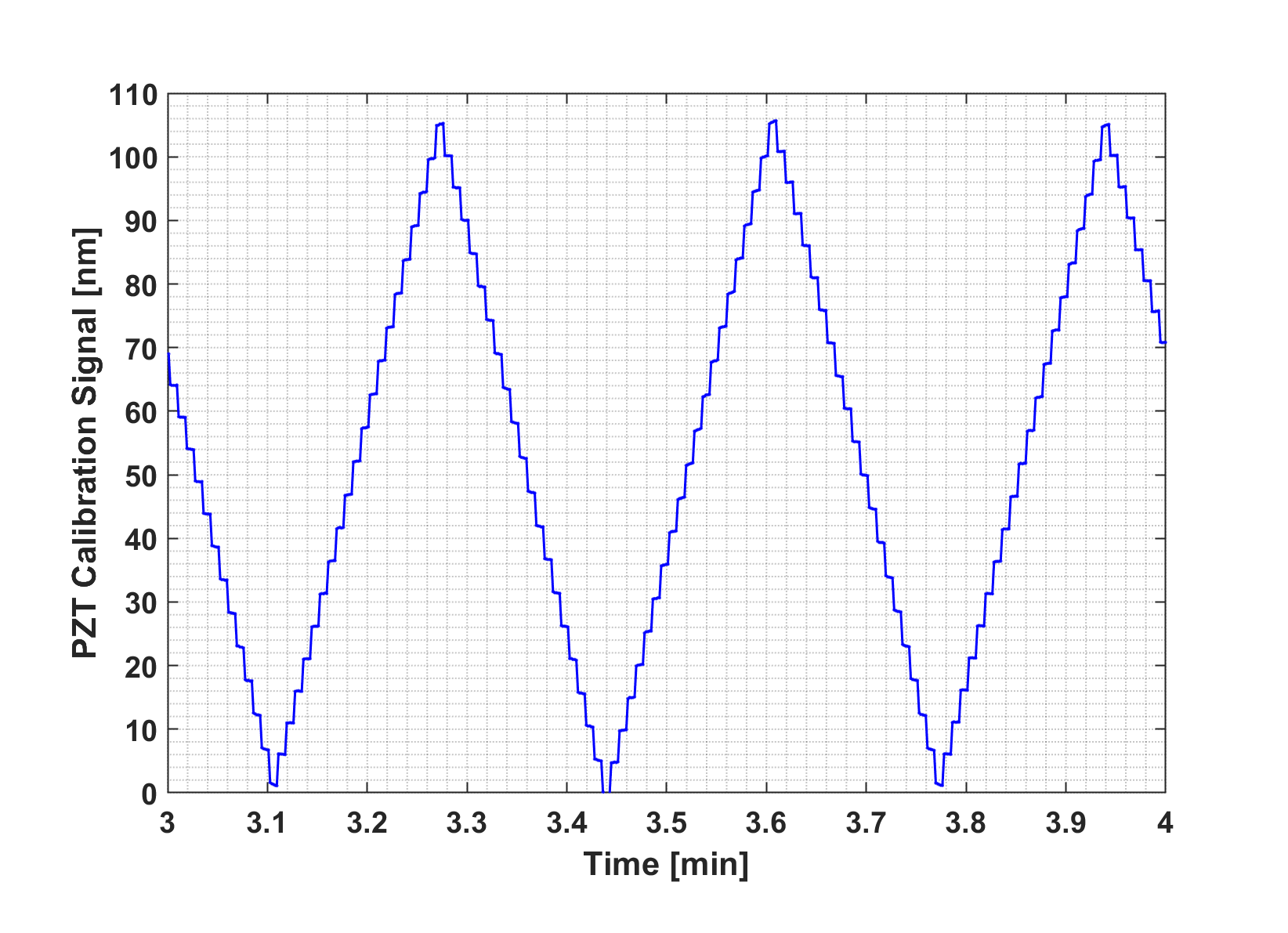}
\caption{The in-orbit calibration of the PZT displacement measured by the M1 Interferometer. Each step is $5~\unm$.}
\label{fig:OP3}
\end{figure}

\begin{figure}
\includegraphics[width=\linewidth]{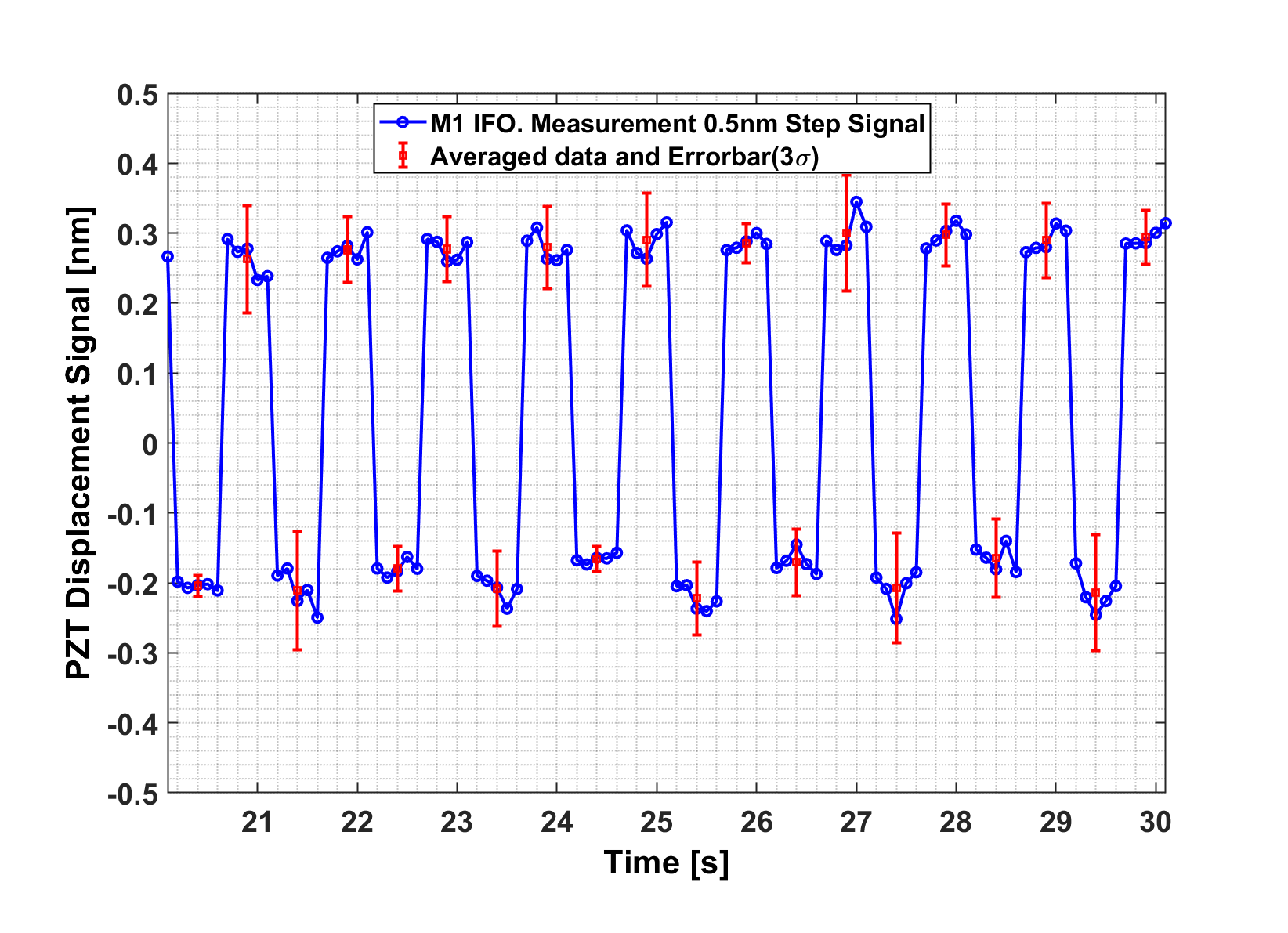}
\caption{The PZT has been commanded to drive a $0.5-\unm$ stepwise motion of the mirror for several steps.}
\label{fig:OP}
\end{figure}

\begin{figure}
\includegraphics[width=\linewidth]{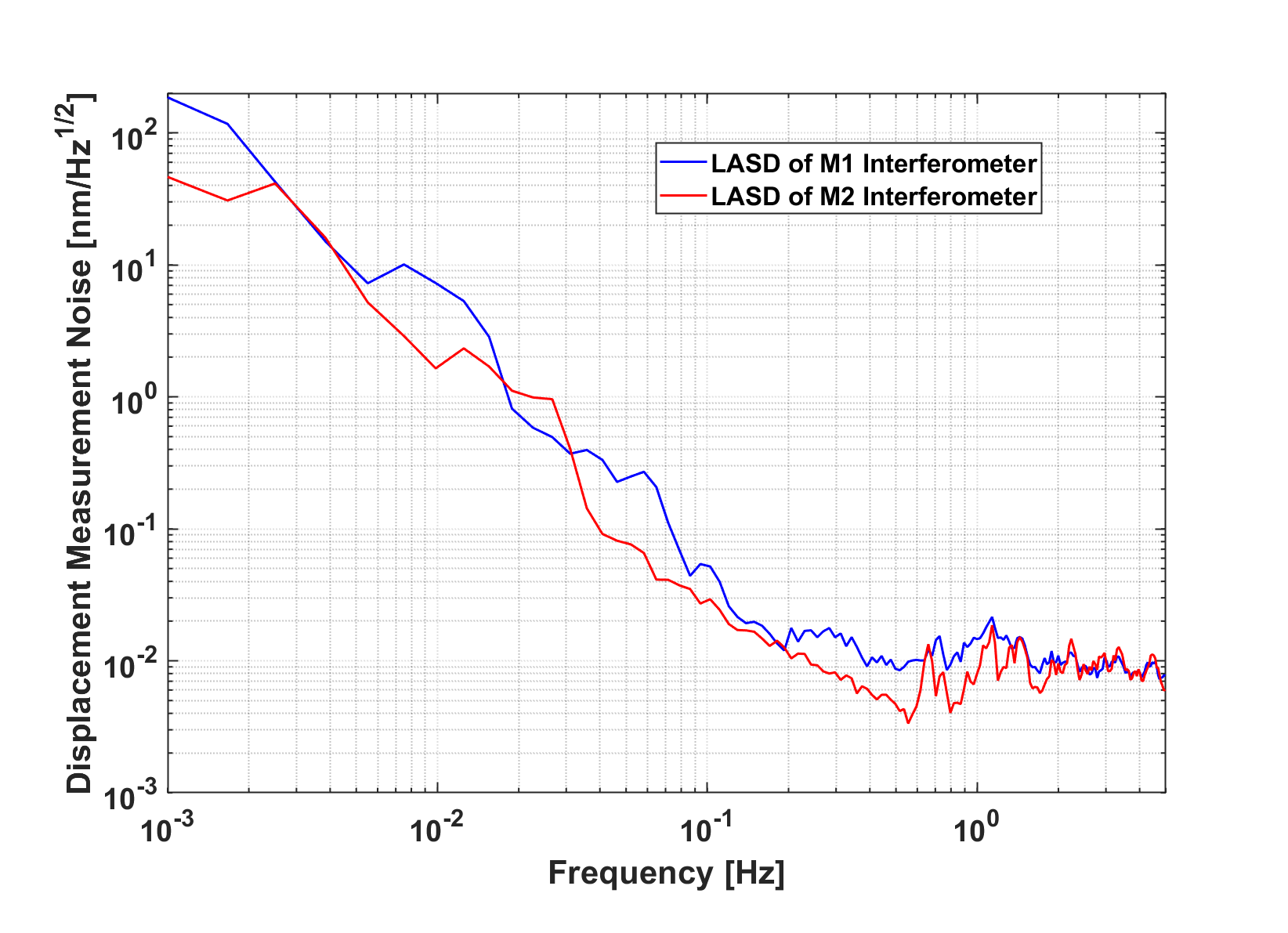}
\caption{The noise spectrums of the M1 and M2 interferometers.}
\label{fig:OP2}
\end{figure}

The in-orbit calibration shows that the piezoelectric constant of the PZT is $(10.40\pm0.39)~\unm/\uV$, while the value measured on the ground is $(10.40\pm0.10)~\unm/\uV$, showing that the on-board PZT is operating as designed.
The in-orbit calibration of the PZT displacement measured by the M1 interferometer is given in Fig. \ref{fig:OP3}.

After calibration, the PZT is used to move the mirror (M70) back and forth with a step of $0.5~\unm$ for several times.
The corresponding displacement measured by the M1 interferometer is shown in Fig. \ref{fig:OP}, in which the long-term drift due to laser frequency shift has been removed from the raw data.
Calculating the standard deviation of five data points for every step, and using the standard method of error analysis, we obtain the measurement resolution of the M1 interferometer to be about $25~\upm$.

The flexure hinge has a natural frequency of $322~\uHz$ and is designed to respond to large disturbances of the satellite, so it can be treated as almost fixed during experiment.
The spectrum density of the measurement noise of the M2 interferometer is shown in Fig.  \ref{fig:OP2}.
The noise level of the M2 interferometer is about $30~\upm/\uHz^{1/2}$ at $0.1~\uHz$.
Notice that the M2 interferometer is less noisy than the M1 interferometer. This is because the pathlength difference of the M2 interferometer ($(0.0\pm3.0)~\umm$) is less than that of the M1 interferometer ($(9.2\pm3.0)~\umm$).

\subsection{Temperature control and CoM measurement}

The temperatures of the key payloads (the inertial sensor and the ORS) are actively controlled to be $20^\circ\uC$ during experiment.
There are 16 high precision platinum-resistance temperature sensors located in the vicinity of the key payloads.
The result from in-orbit experiment shows that a controlled temperature stability of $\pm 3~\umK$ per orbit has been achieved.

Before launching, the CoM mismatch between the TQ-1 satellite and the TM of the inertial sensor is measured and compensated with a precision of $0.1~\umm$. When the satellite is in orbit, the mismatch is determined by changing the attitude variance of the satellite, meanwhile, the linear and angular accelerations of the satellite are measured by the inertial sensor and the gyros \cite{Li2017gp}. The variance of the CoM mismatch has been evaluated with a precision of less than $0.1~\umm$ along each axis.

\section{The implications of TQ-1 for TianQin}\label{sec:discussion}

In this section, we discuss what TQ-1 means for the two core technology requirements of TianQin, i.e., the nongravitational disturbance on the TMs must be reduced to $10^{-15}~\um/\us^{2}/\uHz^{1/2}$ in $10^{-4}\sim1~\uHz$ band, and the measurement noise of laser interferometry must be reduced to $10^{-12}~\um/\uHz^{1/2}$ in $10^{-4}\sim1~\uHz$ band.

To meet the first core technology requirement of TianQin, one has to have high precision inertial sensors with a noise level of $10^{-15}~\um/\us^{2}/\uHz^{1/2}$ in $10^{-4}\sim1~\uHz$ band, high precision micro-Newton thrusters with thrust range of  $1\sim100~\muN$ and thrust noise of $0.1~\muN/\uHz^{1/2}$ in $10^{-4}\sim1~\uHz$ band, and a robust drag-free control loop to reduce the acceleration noises of the TMs to the required level. TQ-1 has demonstrated that the noise level of the inertial sensor is at least $1\times10^{-10}~\um/\us^{2}/\uHz^{1/2}$ at $0.1~\uHz$ and $5\times10^{-11}~\um/\us^{2}/\uHz^{1/2}$ at $0.05~\uHz$, while the estimation on the ground gives a noise level of $5\times10^{-12}~\um/\us^{2}/\uHz^{1/2}$ at $0.1~\uHz$.
These results quantify the gap between the current technology capability and the requirement of TianQin. The performance of the inertial sensor can be improved by using a heavier TM, a larger gap between the TM and the capacitive frame, as well as significantly more stable and strict environmental conditions. TQ-1 has demonstrated that the micro-Newton thrusters can reach the required thrust range of $1\sim60~\muN$, and the estimated thrust noise of $0.3~\muN/\uHz^{1/2}$ is also close to the requirement of TianQin at $0.1~\uHz$. With drag-free control, TQ-1 has measured a residual acceleration noise of about $3\times10^{-9}~\um/\us^{2}/\uHz^{1/2}$ at $0.1~\uHz$, which also quantifies the gap between the current technology capability and the requirement of TianQin.

To meet the second core technology requirement of TianQin, one needs to have the corresponding technologies with optical bench, frequency stabilized laser, weak-light phase locking, pointing control, and so on. The ORS on TQ-1 has achieved a noise level of $30~\upm/\uHz^{1/2}$ at $0.1~\uHz$, which is mainly caused by the laser frequency drifting, the power fluctuation and the electrical noise of phasemeter. The performance of ORS can be improved by using frequency-stabilized and power-stabilized lasers, optimized digital phasemeters and active thermal control with multi-layer thermal shielding. Not only ultra-stable lasers, but also the technologies of weak-light phase locking and laser-beam pointing control will need to be tested in the future.

Stability of the satellite platform is essential in providing the necessary environment where the above core technology goals can be achieved in space. In this regard, TQ-1 has also achieved a stability of temperature control to the level of $\pm3~\umK$ per orbit (97.13 min). In future, more efforts should be put in suppressing any potential sources of noise from the platform.

In summary, the results from the first round of in-orbit experiment show that the performances of all payloads have met (and some are better than) the designed requirements of TQ-1 mission. This successful mission has marked a new milestone in the development of the TianQin project.

\section*{Acknowledgments}

The authors thank the supports from the National Space Administration, the Ministry of Education, the Guangdong Provincial Government and the Zhuhai Municipal Government, of the People's Republic of China. The authors also thank all members of the TianQin collaboration, project partners, and participants at the workshops for the TianQin science mission, for collaboration and helpful discussions. Partial supports for the work also come from the National Natural Science Foundation of China (Grant Nos. 11727814, 11927812, 11655001, 11654004) and the Guangdong Major Project of Basic and Applied Basic Research (Grant No. 2019B030302001).

\section*{References}
\bibliographystyle{iopart-num}
\bibliography{TQ-1}
\end{document}